\documentclass{soups}

\usepackage{balance}       
\usepackage{graphics}      
\usepackage[T1]{fontenc}   
\usepackage{txfonts}
\usepackage{mathptmx}
\usepackage[pdflang={en-US},pdftex]{hyperref}
\usepackage{color}
\usepackage{booktabs}
\usepackage{textcomp}

\usepackage{microtype}        
\usepackage{ccicons}          

\usepackage{todonotes}

\usepackage{xspace}
\usepackage{algpseudocode,algorithm,algorithmicx}
\usepackage{multirow}
\usepackage{tabularx}
\usepackage{amsmath}
\usepackage{xspace}
\usepackage{caption}
\newcommand{\tapmein}{TapMeIn\xspace }

\def\plaintitle{Tap-based User Authentication for Smartwatches\titlenote{Published at \cite{NGUYEN2018174}}}

\def\emptyauthor{}
\def\plainkeywords{Tap-based Authentication; Rhythm-based Authentication; Smartwatch; Shoulder Surfing; Usable Security; Biometrics}

\makeatletter
\def\url@leostyle{%
  \@ifundefined{selectfont}{
    \def\UrlFont{\sf}
  }{
    \def\UrlFont{\small\bf\ttfamily}
  }}
\makeatother
\def\pprw{8.5in}
\def\pprh{11in}

\definecolor{linkColor}{RGB}{6,125,233}
\hypersetup{%
  pdftitle={\plaintitle},
  pdfauthor={\emptyauthor},
  pdfkeywords={\plainkeywords},
  pdfdisplaydoctitle=true, 
  bookmarksnumbered,
  pdfstartview={FitH},
  colorlinks,
  citecolor=black,
  filecolor=black,
  linkcolor=black,
  urlcolor=linkColor,
  breaklinks=true,
  hypertexnames=false
}

\begin{document}

\title{\plaintitle}

\numberofauthors{2}
\author{%
 \alignauthor{Toan Nguyen\\
    \affaddr{New York University}\\
    \email{toan.v.nguyen@nyu.edu}}\\
 \alignauthor{Nasir Memon\\
    \affaddr{New York University}\\
    \email{memon@nyu.edu}}\\
}

\maketitle

\begin{abstract}

  This paper presents \tapmein, an eyes-free, two-factor authentication method for smartwatches.  It allows users to tap a memorable melody (\emph{tap-password}) of their choice anywhere on the touchscreen to unlock their watch. A user is verified based on the tap-password as well as her physiological and behavioral characteristics when tapping. Results from preliminary experiments with 41 participants show that \tapmein could achieve an accuracy of 98.7\% with a False Positive Rate of only 0.98\%. In addition, \tapmein retains its performance in different conditions such as sitting and walking. In terms of speed, \tapmein has an average authentication time of 2 seconds. A user study with the System Usability Scale (SUS) tool suggests that \tapmein has a high usability score.
  
\end{abstract}


\section{Introduction}
\label{sec:tapmein_intro}
Smartwatches are becoming increasingly popular thanks to the seamless experience they offer to consumers~\cite{Gartner}, with their applications ranging from getting notifications, tracking health and fitness, to conducting financial transactions~\cite{ApplePay, SamsungPay}. Recently, smartwatches are being used to conveniently unlock computers~\cite{AppleAutoUnlock, GoogleSmartLock} and even cars~\cite{Ford, Hyundai}. 
However, despite potentially containing sensitive and private information, smartwatches  are not as secure as their counterparts--mobile phones as shown in~\cite{HP, MobileIron}. Specifically, researchers have found that only five out of the ten most popular smartwatch models offer a lock screen method (either a PIN or a Pattern Lock), to protect user information in a stolen device scenario. Further, two in ten devices can be paired with an attacker's smartphone without any difficulty~\cite{HP}. Moreover, protection methods are  often turned off by default and smartwatches usually do not prompt users to enable them (except Apple Watch)~\cite{MobileIron}. 

Even the PIN and Pattern Lock methods, if at all used,  have many weaknesses. They are known to be vulnerable to guessing attacks~\cite{bonneau2012birthday, serwadda2013kids}, shoulder surfing~\cite{wiedenbeck2005passpoints, tari2006comparison, papadopoulos2017illusionpin},  smudge attack~\cite{aviv2010smudge}, and video attack where a whole authentication session could be recorded  with a camera or a device like Google Glass or GoPro~\cite{yue2014my}. From a usability point of view,  authentication with a PIN or a Pattern may suffer from the ``fat-finger problem'' due to the limited size of the smartwatch screen~\cite{siek2005fat}. Moreover, putting biometric sensors like fingerprint scanners and camera for face recognition on smartwatches may be difficult given their small form factors. 

This paper introduces \tapmein, an eyes-free, two-factor authentication method that allows a user to tap a memorable melody or \emph{tap-password} on the smartwatch touchscreen to login. A user is verified based on the correctness of the tap-password as well as  features which depend on the physiological and behavioral characteristics of a user. 

\tapmein offers several desirable features. First, in terms of security,  its two-factor nature makes guessing, smudge and shoulder surfing attacks less relevant. Even in the case where an attacker knows and can repeat the melody of a user, he still needs to pass  behavioral and physiological verification which is significantly more difficult. This protection also applies to video attacks as shown in our evaluation. Second, in terms of usability, its eyes-free feature, allows the user to tap anywhere on the screen, and hence not only solves the fat-finger problem but also enables users to login discreetly and benefits users with visual impairment. In different conditions like sitting and walking, \tapmein achieves performance similar to  that of PIN and Pattern Lock methods.

It is envisioned that \tapmein can be deployed as lock/unlock method or to secure pairing between a watch and a phone. It can also be displayed as an option along the PIN or Pattern for the user to choose based on the context of usage or surroundings. For example, users can choose \tapmein to unlock their watch in a public place where the risk of being observed by someone is high. And when they are at home or alone, they can unlock their watch with Pattern Lock or the regular PIN method.

The main contributions of this paper are summarized as follows.
\begin{itemize}
\item We propose \tapmein, an eyes-free, two-factor authentication  for smartwatches with a touchscreen which provides resilience against guessing, smudge, shoulder surfing and, to some degree, video attacks.
\item We introduce and evaluate a new feature set for tap-based authentication.
\item We present comprehensive evaluation results of \tapmein through a study involving 41 participants, in different contexts (sitting and walking) and in several attack scenarios. 
\item We present a comparative study of several authentication methods on smartwatches, including \tapmein and de-facto PIN and Pattern Lock authentication. The results show that while there is no significant difference in error rate between \tapmein and the other two methods, \tapmein provides a much higher resilience against shoulder surfing and video attacks. In terms of login speed, \tapmein is slightly slower than Pattern Lock but there is no significant difference with 4-digit PIN authentication. PIN and Pattern Lock have been evaluated extensively on smartphones, however, study of their performance on smartwatches is sparse. We provide interesting insights as a baseline for future research on this topic.
\end{itemize}

The rest of this paper is organized as follows: The  threat model and design of \tapmein as well as its modules are presented in Section~\ref{sec:tapmein_system}. Data collection and the performance evaluation results (the first study) are detailed in Section~\ref{sec:tapmein_data_collection} and  Section~\ref{sec:tapmein_evaluation}, respectively. The second study, detailed in Section~\ref{sec:tapmein_comparisons}, presents a comparison between \tapmein and the two methods currently available on smartwatches: PIN and Pattern Lock. Discussion and limitations are described in Section~\ref{sec:tapmein_discussion}. Related work is presented in Section~\ref{sec:tapmein_related_work}. Section~\ref{sec:tapmein_conclusion} concludes this paper.

\section{TapMeIn system design}
\label{sec:tapmein_system}
In this section, we first present the threat model and give an overview of \tapmein. Then details of its modules are described in subsequent subsections. 

\subsection{Threat Model}
\label{subsec:tapmein_threat_model}
We designed \tapmein with a focus on shoulder surfing and video attack threat models. In the shoulder surfing scenario, an attacker obtains a user's tap-password by observing a user unlocking her watch using \tapmein from reasonably close proximity. The observation can be made once or multiple times. In the video attack, a determined attacker goes to a great length to impersonate a user by video-recording an authentication session of the user. He then learns the user's tap-password by watching the video carefully and practicing its performance  then tries to imitate the user. In the next few sections, we present detailed design of \tapmein and a study to evaluate \tapmein performance under these threat models.

\subsection{System overview}
\label{subsec:tapmein_sytem_overview}
\begin{figure*}
\centering
  \includegraphics[width=\textwidth]{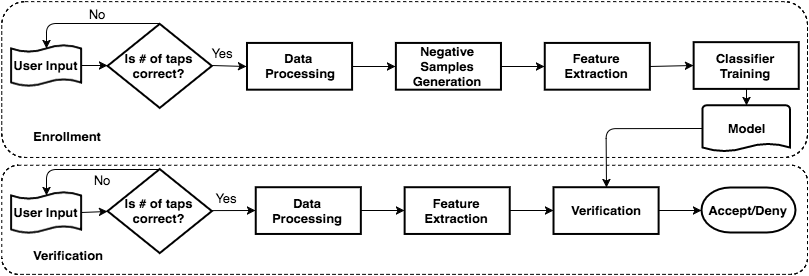}
  \caption{\tapmein system overview}
 \label{fig:tapmein_system_overview}
\end{figure*}

Like any other biometric system, \tapmein consists of two phases: \emph{Enrollment} and \emph{Verification} as depicted in Figure~\ref{fig:tapmein_system_overview}. 
\begin{itemize}
\item \textbf{Enrollment}: In the enrollment phase, a user is asked to choose a melody as her tap-password and requested to provide samples of her tap-password. Each enrollment sample is first checked for its length (number of taps). If the length is not correct, the user is prompted to reenter it. The sample is then passed to a \emph{Data Processing} module for extraction of data used in later steps. At this phase, there are no negative samples for the training of the classifier. We choose an approach to generate synthesized negative samples as detailed in a subsection below. After this step, all samples (enrollment and synthesized negatives) are sent to a \emph{Feature Extraction} module where feature vectors are extracted as inputs to a \emph{Classifier Training} module. The Classifier Training module trains and outputs a model of a binary classifier for the user.

\item \textbf{Verification}: In the verification phase, anyone attempting to unlock the watch needs to enter a tap-password. \tapmein first verifies that the entered tap-password has the correct length. If not, access to the watch will be rejected immediately. Otherwise, the input tap-password also goes through Data Processing and Feature Extraction modules. The resulting feature vector is fed into a \emph{Verification} module, where the user classifier's model is applied to determine whether the entered tap-password is legitimate and access should be granted or not.
\end{itemize}

Details of \tapmein's modules are presented in following subsections.

\subsection{User Input}
\label{subsec:tapmein_user_input}
 A tap is defined by two events: $TOUCH\_DOWN$ is the event when the finger touches the screen and $TOUCH\_UP$ is the event when it is lifted off. In modern touchscreen devices like smartwatches, when a user taps on the screen, multiple sensor data can be acquired from that event. Among these data, \tapmein uses pressure, size of touch and timestamp values.

Let $T=\{T_i\}_1^l$ represent a tap-password of length $l$ in which each tap is: \\

 $T_i = \{TOUCH\_DOWN_i, down\_ts_i, p_i, s_i, TOUCH\_UP_i, up\_ts_i\}$ where:

\begin{itemize}
\item $p_i$: pressure, indicating how hard the finger pressed on the screen, $p_i \in [0 .. 1]$. Pressure values are often different among different users, because each user has a different tapping behavior.
\item $s_i$: size of touch, indicating how much skin of the finger touched the screen, $s_i \in [0 .. 1]$. Size value depends on both user's physiological and behavioral characteristics. For example, users with bigger fingers might be likely to touch harder on the screen, which results in higher pressure and size values.
\item $down\_ts_i$ and  $up\_ts_i$: the timestamps when the finger touch down and are lifted off the screen
\end{itemize}

\subsection{Data Processing}
\label{subsec:tapmein_data_processing}
This module prepares data for feature extraction.  A melody is comprised of rhythms where each rhythm can be represented by a down and an up duration of each tap. The down duration indicates how long the finger touches on the screen in a tap and the up duration indicates the time between two consecutive taps. The sequence of up and down duration makes up the melody. Even if two users choose the same melody, their interpretations of the melody will be different and that difference will be reflected in their up and down duration sequences. The duration sequence is extracted as follows.

For each tap $T_i$, let $\Delta_i^d$, $\Delta_i^u$ be its down and up duration respectively. Then:
\begin{equation}
\Delta_i^d = up\_ts_i - down\_ts_i; \quad
\Delta_i^u = down\_ts_{i+1} - up\_ts_i
\end{equation}

Note that the last tap does not have the up duration $\Delta_u^l$. The timestamps data are removed, and the tap-password at the end of this process is as follows: 

$T = \{{\bar{T}}_i\}_1^l$ where ${\bar{T}}_i = \{p_i, s_i, \Delta_i^d, \Delta_i^u\}$. In other words, $T$  consists of four timeseries of pressure, size of touch, down and up duration.

After pre-processing, the tap-password is sent to the \emph{Feature Extraction} module.

\subsection{Feature Extraction}
\label{subsec:tapmein_feature_extraction}
This module takes processed tap-password $T=\{{\bar{T}}_i\}_1^N$ from the Data Processing module and extracts a vector containing features used in \tapmein. We extract both time-domain and frequency-domain features.

\textbf{Time-domain features}. 
First, all values in the four timeseries are added to the feature vector. Then, we extract statistical features from each timeseries including \emph{min, max, mean and variance}.

\textbf{Frequency-domain features}.
We transform each timeseries into its frequency representation using Fast Fourier Transform (FFT). We then calculate magnitudes of FFT components. Statistical measures of these magnitudes are added to the feature vector including \emph{min, max, mean and variance} values. Finally, energy, computed as sum of all magnitudes of each timeseries, is added to the feature vector.

The output feature vector is then passed to a \emph{Classifier Training} module in the enrollment phase or to a \emph{Verification} module in the verification phase.

\subsection{Negative Samples Generation}
\label{subsec:tapmein_generating_negative_samples}
Two challenges that  have to be addressed to devise a good classifier model for \tapmein are the lack of negative or imitation samples and a small sample set for classifier training. This is because, first, we usually do not have impostor's samples in the enrollment phase and, second, requiring a user to provide many samples for training is not reasonable and may harm the usability of the system. As these are common problems, several approaches have been applied in the literature to address them. 

The first approach is to use a distance-based classifier which determines the legitimacy of an entered password by comparing its distance to enrollment samples with a predetermined threshold~\cite{wobbrock2009tapsongs, lin2011rhythmlink, marques2013under, van2014finger, sae2012biometric, nguyen2017draw}. The distance function can be Hamming~\cite{marques2013under}, Euclidean~\cite{lin2011rhythmlink}, or Dynamic Time Warping (DTW)~\cite{van2014finger, sae2012biometric, nguyen2017draw}. However, this approach usually does not give the desired performance for tap-based authentication: either a low accuracy (i.e., 83.2\% as reported in~\cite{wobbrock2009tapsongs}) or a high false positive rate (i.e., 5 false positives out of every 30 impostor's trials = 16.67\%  as reported in~\cite{marques2013under}). 

The second approach is to use a one-class classifier, which is trained on a data set containing only one type of samples, i.e., legitimate samples, but can be used to classify both legitimate and illegitimate samples at verification phase~\cite{shahzad2013secure}. This approach, unfortunately, also requires a significant number of enrollment samples for training. 

A third approach is to preload negative samples onto the device or download them from a server at the time of classifier training~\cite{chen2015your}. This proposal has a significant drawback, which is, besides the fact that we do not know in advance what kind of tap-password or melody a user will choose, totally random negative samples usually bias the classifier. For example, the classifier performs well on training because randomly generated negatives are  usually distinguishable from legitimate samples, but performs poorly for verification when real imitation samples from the impostors are used, i.e., samples from a shoulder surfing attack. We have observed this problem in our experiments. 

Finally, another plausible approach is to generate synthesized negative samples, which can help to address both challenges: producing negative samples as well as increasing the training set size. We adopt this approach for \tapmein. We tried different ways of generating negative samples and observed \tapmein performance in each case. We found that generating completely random samples does not work as the classifier can easily separate two classes in the training set, but fails to classify real impostor samples. 
Our approach for generating negative samples is as follows. First, we collect many tap-passwords from diverse subjects to build \emph{distributions} of each feature in tap-passwords. Specifically, we will have a distribution for pressure, a distribution for size of touch, a distribution for down duration and a distribution for up duration. We only need to ship representations of these distributions (min, max, mean, standard deviation) with \tapmein, which are negligible in size. To generate a negative sample for a user, we draw from each distribution a random value for each tap feature (pressure, size, down and up duration). This process stops when the number of taps equals to the length of user's tap-password. The procedure can be repeated many times to generate a desired number of samples. The set consisting of the user's enrollment and synthesized samples is then passed to Feature Extraction and eventually to the Classifier Training module to build a model for the user's tap-password.

\subsection{Classifier Training}
\label{subsec:tapmein_classifier_training}
This module trains a binary classifier for each user using the feature vectors extracted from the training set of the user. We choose Support Vector Machines (SVM)~\cite{cortes1995support} and Random Forests~\cite{breiman2001random} as our classifiers.  In a binary classification problem like ours, SVM builds a hyperplane to separate feature space such that it maximizes the margin between the two classes. Random Forests is an ensemble learning method that constructs a forest, which is a combination of many decision trees, at training time. The forest is used to classify an input vector by inputting the vector to every tree in the forest. Each tree outputs a classification and the forest chooses the class with the most ``votes" among the trees as the final prediction. To tune parameters for each classifier, we run a grid search with different parameter sets on the training data and pick the set that returns the best performance, i.e., accuracy.

\subsection{Verification}
\label{subsec:tapmein_verification}
At verification time, a candidate user input, after passing the length test, goes through the same Data Processing and Feature Extraction modules. A feature vector is extracted and fed into the Verification module, which tests it using the classifier model of the user and determines whether this input is legitimate or not. 

\section{Data Collection}
\label{sec:tapmein_data_collection}
A study was conducted to evaluate the performance of \tapmein in terms of security and usability. Because smart watches are often used on the move, i.e., walking or running,  the study was designed to explore the performance of \tapmein in different conditions. Specifically, when users are sitting and walking. We implemented \tapmein on a Samsung Gear Live Smartwatch running Android Wear OS. In the rest of this section, the study design and data collection procedure are presented. Note that all studies presented in this paper were approved by the institutional review board at our university.

\subsection{Study Design}
\label{subsec:tapmein_study_design}
The study was conducted using a repeated measures design. The independent variable was \emph{Condition} which has two levels: \emph{Sitting} and \emph{Walking}. The dependent variables were performance figures (accuracy, error rate and authentication time).  Resilience against shoulder surfing and video attack were also evaluated.

\subsection{Participants}
41 participants (females=13) with an average age of 27.5 (range: 18 to 57) were recruited through email and Facebook postings. All participants had university education and were smartphone users and  eight had experience with smartwatches before this study. 31 participants participated as users and 10 as attackers in our experiments.

\subsection{Procedure}
\label{subsec:tapmein_procedure}
All participants performed the authentication task on the same smartwatch and the same ambient conditions. Upon their arrival to the lab, each user was introduced to the study and was given the opportunity to operate the smartwatch to get used to it. She was instructed to wear the watch on the hand she usually wears one. She was then introduced to \tapmein and was asked to choose a melody as her tap-password. She would start entering her tap-password after pressing a login button. Once she finished tapping, she pressed an \emph{enter} button at the bottom right corner of the screen. The device processed her tap-password, returned a login result and displayed a new screen for another login session. She was allowed to practice with \tapmein until she felt comfortable with it and was able to successfully log in using this method at least five times for each condition, sitting and walking. Data in this phase was not collected.

The authentication phase started with an enrollment of the user. She was asked to tap her melody three times when sitting for training. The melody was the same as in the practice phase to minimize learning effect. She was then asked to login ten times while sitting and ten more times while walking around the room. For counterbalancing, half the users were randomly chosen to perform login while sitting first and the other half was required to perform login while walking first. Once the authentication phase was completed, the users answered a survey on their perception of the security and usability of \tapmein. Overall, the study lasted about an hour for each user.

\subsection{Data Collection for Shoulder Surfing and Video Attacks}
To simulate shoulder surfing and video attacks, we video-recorded authentication sessions of each user. An experiment conducter playing the role of an attacker stood behind each user's shoulder, on the side where the user's hand did not obscure the watch screen. This attacker recorded, using a smartphone, several authentication sessions of the user as she was seated. For security analysis, the set of videos recorded were cut down to one successful authentication for each user. The tap-password input process was clearly observed in each video with no obstruction of user's hand or shoulder. This setup favored the attackers. However, we wanted to evaluate \tapmein in such situations.

10 participants, who were not users, were asked to simulate impostors/attackers. They went through the same introduction phase as users to get comfortable with \tapmein and the purpose of this study. The attackers were placed in our lab with the same smartwatch and a computer with 21-inch monitor. All videos were played with sound and were recorded in 1080p using a Nexus 4 phone. Each attacker imitated all 31 users in three attacks as follows. The entire study took about three hours for each attacker, so the attacker was required to take several breaks during the process.

\begin{itemize}
\item \textbf{Attack-I: One-time shoulder surfing}: The attacker watched a user's video once and was allowed three trials to unlock the watch using the tap-password that he collected.
\item \textbf{Attack-II: Two times shoulder surfing}: The attacker watched the user's video one more time and was allowed three more trials to unlock the watch.
\item \textbf{Attack-III: Video Attack}: Finally, the attacker was allowed to watch, rewind and replay the user's video as long as he wanted and was allowed three more trials to unlock the watch.
\end{itemize}

\section{Performance evaluation results}
\label{sec:tapmein_evaluation}
In this section, we report performance evaluation results of \tapmein based on the collected data set.
We use False Positive Rate (FPR), False Negative Rate (FNR) and Equal Error Rate (EER) to report performance in different experiments. FPR shows how often a system falsely accepts attackers, and FRR indicates how often it wrongly rejects legitimate users. EER is a common metric in biometric systems. EER is the point where FPR and FNR are equal. Thus, EER balances between usability and security. 

In terms of login speed, authentication time was used for measurement. However, the time it takes to authenticate an input is small, thus, authentication time mostly indicates input time.

There were 31 users, each providing 13 samples while sitting (3 enrollments + 10 logins) and 10 samples while walking, resulting in a data set of 31 $\times$ 13 = 403 legitimate samples in the sitting condition and 31 $\times$ 10 = 310 legitimate samples in the walking condition. There were three attacks, which each had 10 attackers and each attacker provided 3 imitation samples for each user, resulted in a set of 31 (users) $\times$ 10 (attackers) $\times$ 3 (samples) = 930 samples for each attack. Thus, our analysis was based on the total of 713 legitimate and 2790 imitation samples.

Evaluation was performed for each user as follows. Recall that each user needed to provide a small number of samples in the enrollment phase for classifier training, which was denoted as $n$. For each user, we randomly chose her $n$ samples to add to her $TrainingSet$. We then generated $5n$ negative samples and added them to $TrainingSet$. We found that classification performance remained stable when the number of generated negatives reached  $5n$ and beyond, so $5n$ samples were enough. The classifier of each user was trained with her $TrainingSet$. The remaining samples of the user in a test condition (sitting, walking, or both, depends on each test we run) and samples from attackers were used for testing. To avoid any bias such as randomness in negative samples, we did this evaluation 30 times for each user, and the results were the averaged over these 30 times. The system performance was then averaged over all users.

All evaluations were implemented using \emph{scikit-learn}\footnote{https://scikit-learn.org}, a popular Python library for machine learning. For tuning classifiers' parameters, we ran a grid search for each classifier with different parameter sets on the data. We selected the set that resulted in best performances. 

\begin{figure}[!t]
\centering
  \includegraphics[width=\columnwidth]{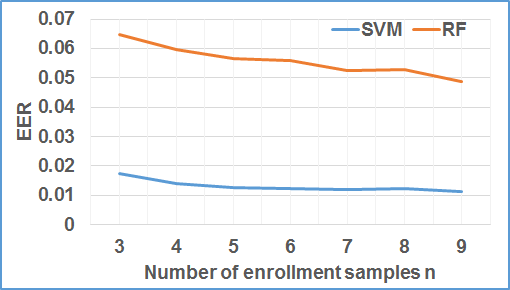}
  \caption{EERs of SVM and RF and the effect of enrollment set size}
 \label{fig:tapmein_training_size}
\end{figure}

\subsection{Performance of different classifiers and the effect of enrollment set size}
We report the impact of different classifiers and enrollment set sizes on the classification error rate. For this evaluation, the remaining samples of a user including all sitting and walking samples, and samples from every other user were used for testing the classifier of the user. The size of the training set was changed by varying $n$ and results are reported in Figure~\ref{fig:tapmein_training_size}. First, we can see that SVM outperformed RF in all different enrollment set sizes. Second, \tapmein achieved stable low error rate when $n>=5$. A smaller $n$ means shorter enrollment time. Thus, we chose $n = 5$ and used SVM for subsequent evaluations. 

\subsection{Features ranking and selection}

\begin{figure}[!t]
\centering
  \includegraphics[width=\columnwidth]{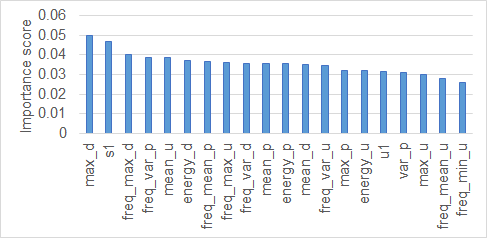}
  \caption{Top 20 features ranked by importance scores.}
 \label{fig:tapmein_feature_importance}
\end{figure}

\begin{table*}[!t]
\centering
\caption{Performance comparison between \tapmein and previous works.}
\label{tbl:tapmein_comparison_with_previous_works}
\begin{tabular}{l|l|l|l|l}
                                        & Accuracy & FPR - Random Attack & FPR-Shoulder Surfing & \# subjects \\ \hline
TapSongs \cite{wobbrock2009tapsongs}   & 83.2\%   & NA                  & 10.7\%               & 10          \\ \hline
Marques et. al \cite{marques2013under} & NA       & NA                  & 16.67\%              & 30          \\ \hline
RhythmLink \cite{lin2011rhythmlink}    & 92\%     & 3\%                 & NA                   & 8           \\ \hline
RhyAuth \cite{chen2015your}            & 95.8\%   & 0.7\%               & NA                   & 32          \\ \hline
\tapmein~\textbf{{[}This work{]}}                & 98.7     & 0.98\%              & 2.3\%                & 41         
\end{tabular}
\end{table*}

To understand the importance of individual feature and to find the most influential ones, we did feature ranking and selection as follows. We trained a Random Forest for each user and configured it to return importance score of each feature. This task was done 30 times for each user with $n=5$ randomly chosen enrollment samples each time. We then ranked the top 20 features for the user based on frequency and importance score of features. Figure~\ref{fig:tapmein_feature_importance} depicted the top 20 features among all users ranked by importance scores. Here, \emph{`p'} is short for pressure, \emph{`s'} is size of touch, \emph{`d'} is down duration and \emph{`u'} is up duration. We can see that there are only two raw tap values, \emph{s1} and \emph{u1}, which are the size and up duration of the first tap. The rest are statistical values of all taps in both time-domain and frequency-domain. This suggests that most raw tap values are not good features. Also, there is a good mix of time-domain and frequency-domain features, which confirms the effectiveness of our proposed features.

\begin{table}[!t]
\centering
\caption{Error rates of \tapmein in two conditions}
\label{tbl:tapmein_different_conditions_performance}
\begin{tabular}{l|l|l}
        & FPR                     & FRR   \\ \hline
Sitting & \multirow{2}{*}{0.98\%} & 5.3\% \\ \cline{1-1} \cline{3-3} 
Walking &                         & 9.1\%
\end{tabular}
\end{table}

\subsection{Performance in different conditions}
In this section, we evaluate \tapmein's performance in different conditions, specifically, sitting and walking. First, we trained a classifier for each user using $n=5$ samples randomly selected from user's samples set. User samples from each condition (sitting/walking) and samples of other users were used for testing. We configured the classifier to return classification label (positive/negative) instead of returning probability of each class. This way, the comparison of error rates between the two conditions is fair because the same threshold was applied. The results are reported in Table~\ref{tbl:tapmein_different_conditions_performance}. Negative samples were the same for the two conditions, but legitimate samples were drawn from the data set of each condition. Therefore, we can see that FPR is the same in two conditions. FRR, however, are different between two conditions due to different numbers of false negatives in each condition. A paired-samples t-test was conducted to compare FRR for sitting and walking conditions. There was a significant difference in the FRR in two conditions  ($t(30)=2.434$, $p < 0.0001$, with $CI=[0.006, 0.069]$). Specifically, the results show that FRR increased when users were walking during authentication compared to sitting. This is expected as when people are walking, their hands are not as stable, and that may cause more errors in their tapping behavior. Nevertheless, being rejected 5.3 and 9.1 times out of every 100 trials when sitting and walking, respectively, could be acceptable, especially considering the security enhancement that \tapmein offers to the users.

\subsection{Security evaluation}
The overall performance of \tapmein against attacks in the chosen threat model is reported in Table~\ref{tbl:tapmein_overall_performance}, where legitimate samples from both sitting and walking, combined with imitation samples from each attack were used for testing. It can be seen that \tapmein achieves good performance in terms of security. Specifically, the EER is as low as 1.3\% in a Random guessing attack, and only 2.3\% in Attack-I (one-time shoulder surfing). It increases to 3.5\% in Attack-II when the attackers had another observation of user's authentication session. In the video attack scenario (Attack-III), despite gaining significant knowledge about user's tap-password by watching user's video carefully over and over again, the attackers achieve only 4.1\% success rate. These results show that \tapmein is potentially resilient against shoulder surfing and video attacks.

\begin{table}[!t]
\centering
\caption{Performance of \tapmein under attacks}
\label{tbl:tapmein_overall_performance}
\begin{tabular}{ll|l}
                                                       &            & EER   \\ \hline
\multicolumn{2}{l|}{Random Attack}                                   & 0.013 \\ \hline
\multicolumn{1}{l|}{\multirow{2}{*}{Shoulder Surfing}} & Attack-I   & 0.023 \\ \cline{2-3} 
\multicolumn{1}{l|}{}                                  & Attack-II  & 0.035 \\ \hline
\multicolumn{1}{l|}{Video attack}                      & Attack-III & 0.041
\end{tabular}
\end{table}

\subsection{Comparison with previous work}

In this section, we compare the performance of \tapmein with previous works on tap-based authentication, which were primarily proposed for smartphones. \tapmein is the first tap-based authentication evaluated for smartwatches.

Several tap-based authentication approaches have been introduced in the literature. Wobbrock et al. proposed TapSongs, which authenticated a user based on her tapping rhythms input from a binary sensor \cite{wobbrock2009tapsongs}. TapSongs only used up--down durations and was based on timing comparison for verification. In \cite{marques2013under}, Marques et al. proposed to turn up-down duration sequence into a binary string and used Hamming distance comparison for user verification. In \cite{lin2011rhythmlink}, Lin et al. proposed to let users pair their different devices by a tap-based password on both devices. They applied Euclidean distance for tap-password verification. In \cite{chen2015your}, Chen et al. proposed RhyAuth, a user authentication for mobile devices, in which, users can tap or slide a rhythmic password on the touchscreen to unlock their phone. RhyAuth also utilized tapping behavior as a second authentication factor. Table~\ref{tbl:tapmein_comparison_with_previous_works} provides a performance comparison between \tapmein in these works. All figures were taken from corresponding papers in similar evaluation scenarios (random attack or shoulder surfing). We can see that \tapmein outperforms previous work in terms of accuracy and the resilience against random attack as well as shoulder surfing. RhyAuth achieved slightly lower FPR of 0.7\% in random attack compared to 0.98\% of \tapmein. However, \tapmein achieved higher accuracy (98.7\% compared to 95.8\% of RhyAuth), which indicates \tapmein has lower false rejection rate.

In summary, our work presents several advances and contributions over previous approaches. First, unlike previous work that only used time-domain features, we propose a new set of features which combines both time-domain and frequency-domain features for verification of tap-based passwords. This new feature set has resulted in better performances as demonstrated above. We also evaluate the importance of each feature and provide the insight into their effectiveness. This was not done in previous approaches. Second, we present and demonstrate a plausible approach to generate synthesized negative samples for tap-based authentication, which has important implications. It helps eliminate a popular assumption made by similar approaches like RhyAuth that imitating samples from attackers are known and needed for training the classifier. Our negative sample generation approach helps train a better classifier for TapMeIn and makes deploying it in real-world applications and devices much easier as it requires no assumption about user's choice of tap-password. Last but not least, we present comprehensive user studies to evaluate both security and usability of TapMeIn as well as its comparison with state-of-the-art PIN and Pattern Lock method (will be presented in the subsequent section). Interesting insights from our studies will serve as a baseline and inspire future research.

\subsection{Authentication time}
\begin{figure}[!t]
\centering
  \includegraphics[width=\columnwidth]{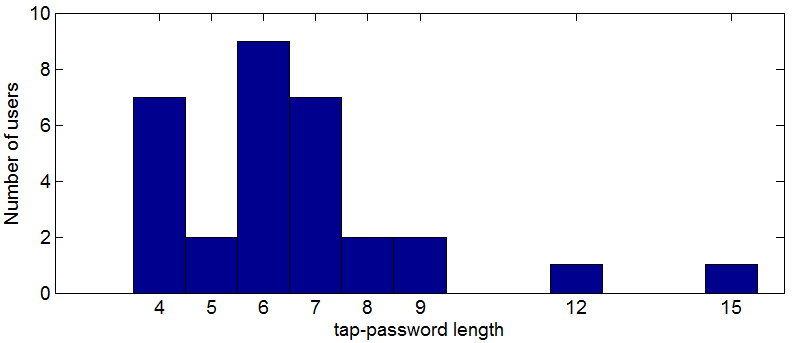}
  \caption{Frequencies of tap-password's lengths among 31 users}
 \label{fig:tapmein_taplength}
\end{figure}

\begin{figure}[!t]
\centering
  \includegraphics[width=\columnwidth]{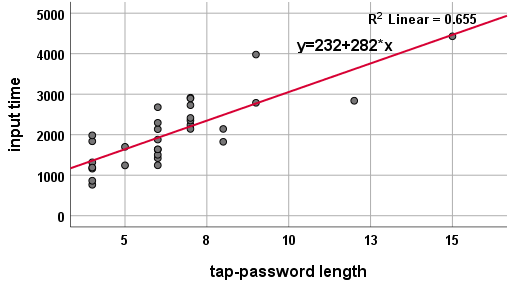}
  \caption{Regression model built to explain input time based on tap-password length}
 \label{fig:tapmein_input_time}
\end{figure}

Authentication time includes the time it takes a user to enter her tap-password (\emph{input time}) and the time it takes to classify the user input (\emph{verification time}). Once the classifier is trained, verifying a user input is fast: verification time is only 24 ms on average in \tapmein as observed in our implementation on a Samsung Gear Live Smartwatch. Thus, the authentication time mostly indicates input time, which depends on the length of the user's tap-password as well as a user's interpretation of a melody. Figure~\ref{fig:tapmein_taplength} details the frequencies of tap-password's lengths chosen by 31 users. It can be seen that the majority of the users chose a tap-password with less than 10 taps. We fit a linear model using the tap-password length to explain the average input time and present it in Figure~\ref{fig:tapmein_input_time}. The tap-password length significantly predicted the average input time ($b = 282$, $t(29) = 7.42$, $p < 0.0001$). The overall model also predicted the average input time very well (adjusted $R^2= 0.655$, $F(1, 29) = 55.12$, $p < 0.0001$). 

In general, it clearly shows that the more taps a tap-password has, the more time is needed to enter it. From Figure~\ref{fig:tapmein_input_time}, we can also see that the average input time of tap-passwords with the same length also varies. This is due to different choices of melodies and different interpretations of the users. The users who chose melodies with rapid notes were likely to tap faster, resulting in shorter input time. Overall, the average input time over all users was 2069 ms (sd = 834 ms). Input time when sitting (mean=2060 ms, sd=856 ms) and walking (mean=2099 ms, sd=805 ms) show no significant difference. Paired-samples t-test was run on average input times of all users in two conditions and showed no statistical significance effect of \emph{Condition} on input time ($t(30) = 0.954$,  $p=0.348$, Cohen's $d=0.18$); thus, we cannot conclude that walking or sitting would increase or decrease authentication time. 

\begin{figure}[!t]
\centering
  \includegraphics[width=\columnwidth]{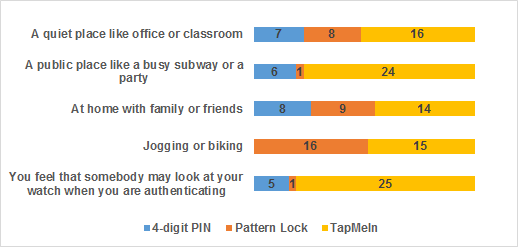}
  \caption{Usecases of \tapmein as rated by the users}
 \label{fig:tapmein_usecases}
\end{figure}

\subsection{Users' perceived usability}
\label{subsec:tapmein_usability}
At the end of the study, each user answered a survey consisting of three parts. The first part was 10 questions from the System Usability Scale (SUS) tool, which is a fast and reliable method for gathering subjective assessments about the usability of a system~\cite{brooke1996sus}. SUS is widely used in the literature to evaluate system usability ~\cite{bangor2008empirical}~\cite{brooke1996sus}. SUS consists of 10 questions,  and each can be answered by a Likert-scale ranging from 1 (Strongly Disagree) to 5 (Strongly Agree). SUS score is a value in the range of 0--100, where a higher value indicates a more usable system. SUS score of a system is calculated by averaging individual scores of all users. Overall, \tapmein achieves a SUS score of 83, which, compared to the standard average score of 68~\cite{SUS}, suggests that \tapmein is highly usable.

The second part of our survey consisted of four following Likert-scale questions designed to gather a user's perceived security and speed of \tapmein. Note that these questions were alternately positively and negatively phrased to avoid bias in user responses.
\begin{itemize}
\item Q11: I think \tapmein is very secure (1-Strongly Disagree, 5-Strongly Agree)
\item Q12: I think guessing the secret rhythmic tap-password is easy (1-Strongly Disagree, 5-Strongly Agree)
\item Q13: I think \tapmein is very fast (1-Strongly Disagree, 5-Strongly Agree)
\item Q14: I will not use \tapmein in the future if it becomes a product (1-Strongly Disagree, 5-Strongly Agree)
\end{itemize}

\begin{table}[!t]
\centering
\caption{User's ratings of four questions in the second part of the survey}
\label{tbl:tapmein_usability_scores}
\begin{tabular}{|l|l|l|}
\hline
                        & mean & sd  \\ \hline
Q11-Security rating     & 4.5  & 0.74 \\ \hline
Q12-Guessability rating & 1.56 & 0.71 \\ \hline
Q13-Speed rating        & 4.03 & 0.73 \\ \hline
Q14-Unlikelihood of future usage        & 1.59 & 0.87 \\ \hline
\end{tabular}
\end{table}

Table~\ref{tbl:tapmein_usability_scores} shows the average ratings of four questions. We can see that, most users perceived \tapmein as secure and fast (Q11 and Q13 have high mean values), and they rated that it would not be easy to guess someone's tap-password. The users also expressed that they would like to use \tapmein in the future. These results confirm that \tapmein is usable and is favorable for the users.

In the third part of the survey, we asked the users about use cases of \tapmein. The question was ``In which situation you would use \tapmein to unlock your watch, consider that you have PIN, Pattern Lock, and \tapmein as available options," with following choices and a text box for another suggestion.
\begin{itemize}
\item S1: A quiet place like office or classroom
\item S2: A public place like a busy subway or a party
\item S3: At home with family or friends
\item S4: Jogging or biking
\item S5: You feel that somebody may look at your watch when you are authenticating
\end{itemize}

The results are depicted in Figure~\ref{fig:tapmein_usecases}. Interestingly, \tapmein was chosen by most users in a situation where shoulder surfing threat is stronger (S2 and S5). This was also confirmed by our informal interview with users after the study. The users also suggested that \tapmein can be deployed as a secure method alongside PIN or Pattern Lock, which can be invoked when they need a higher security level based on their surrounding.

\section{Comparison to PIN and Pattern Lock Authentication}
\label{sec:tapmein_comparisons}
In this section, a second study is presented with the  goal  to compare \tapmein with two de-facto methods currently available on smartwatches, PIN and Pattern Lock, in terms of error rate, security and authentication time. In the PIN and Pattern Lock methods,  error rate is defined as the ratio of the number of user mistakes divided by the total number of user trials. In \tapmein, the error rate is the FRR which includes user mistake as well as classifier errors.

\subsection{Study Design}
This study was conducted using a repeated measures factorial design. The independent variables were \emph{Method} and \emph{Condition}. \emph{Method} has three levels: \emph{PIN}, \emph{Pattern Lock} and \emph{\tapmein}. \emph{Condition} has two levels: \emph{Sitting} and \emph{Walking}. The dependent variables were \emph{error rate} and \emph{authentication time}. This study was designed and carried out in the same procedure and apparatus as in Study-I for fairness of comparison.

\begin{figure}[!t]
\centering
  \includegraphics[width=\columnwidth]{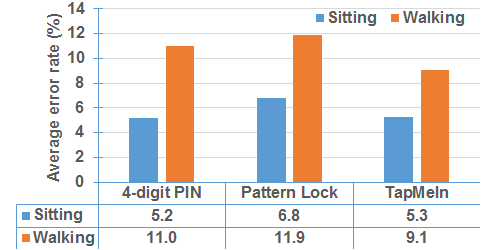}
  \caption{Average \emph{error rate} of three methods in two conditions}
 \label{fig:tapmein_error_rate}
\end{figure}

\subsection{Procedure}
We implemented PIN and Pattern Lock authentication on the same Samsung Gear Live Smartwatch which allowed us to capture the authentication time of each login session from each method. Participants, which were 31 users from Study-I, went through the similar procedure as in Study-I. Specifically, each user was asked to choose a 4-digit PIN for the PIN method and a pattern with length $>= 4$ for the Pattern Lock method. The users were allowed to practice unlocking with the two methods until they felt comfortable with both methods and were able to successfully unlock the watch at least five times using each method in each condition.

In the authentication phase, each user was asked to unlock the watch using each method ten times while sitting and ten times while walking. Counterbalancing was conducted by a Latin square design. For security evaluation, an experimenter recorded a video of a successful authentication session of each user for each method in sitting condition. This study lasted about half an hour for each user.

10 attackers from Study-I also participated as attackers in this study and went through the same process to simulate the three attacks described earlier. The difference is an attacker stopped whenever they got the correct PIN or pattern of the user.

\begin{figure}[!t]
\centering
  \includegraphics[width=\columnwidth]{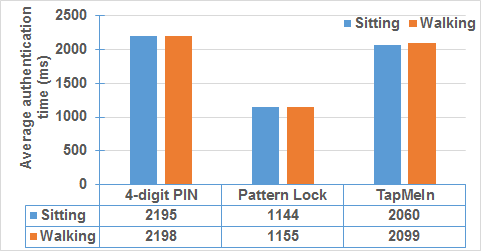}
  \caption{Average \emph{authentication time} of three methods in two conditions}
 \label{fig:tapmein_authentication_time}
\end{figure}

\subsection{Results}
For each method (PIN and Pattern Lock), there were 31 (users) $\times$ 10  (unlocks per condition) = 310 unlocks in sitting condition and 310 unlocks in walking condition. Error rate and authentication time in these two conditions of the three methods are detailed in Figure~\ref{fig:tapmein_error_rate} and Figure~\ref{fig:tapmein_authentication_time}. Note that figures for \tapmein were taken from Study-I. With two-way repeated measures ANOVA, we found significant effect of \emph{Condition} (Sitting/Walking) on the \emph{error rate} ($F(1,30)=12.70$, $p=0.001$, partial ${\eta}^2=0.297$),  but no significant main effect of \emph{Method} (PIN/Pattern Lock/\tapmein) on \emph{error rate} ($F(1.65,49.5)=0.596$,  $p=0.524$, partial ${\eta}^2=0.019$).

Running two-way repeated measures ANOVA on  average authentication times in two conditions in three methods, we found significant  effects of \emph{Method} on \emph{authentication time} ($F(2,60)=28.66$, $p<0.001$, partial $\eta^2=0.49$), but no significant effect of \emph{Condition} on \emph{authentication time} ($F(1,30)=0.064$, $p=0.80$, partial $\eta^2=0.002$). Pairwise post hoc tests using Bonferroni correction showed significant difference in \emph{authentication time} between \tapmein (mean=2079, sd=148) and Pattern Lock (mean=1150, sd=80) ($p<0.001$), but no significant difference between \tapmein and PIN (mean=2180, sd=85) ($p>0.05$). There was also significant difference in \emph{authentication time} between Pattern Lock (mean=1150, sd=80) and PIN (mean=2197, sd=87) ($p<0.001$). These results suggest that, in terms of authentication time, \tapmein is slower than Pattern Lock.

In terms of shoulder surfing resilience, attackers achieved success rates of 93.75\% and 96.88\% on PIN and Pattern Lock method, respectively, in Attack-I and 100\% in Attack-II for both methods. None of them needed to do video attack (Attack-III) as they all succeeded after Attack-II. Compared to attackers' success rate of 2.3\% (Attack-I) and 3.5\% (Attack-II) in \tapmein, we can see that \tapmein is far more resilient against shoulder surfing than PIN and Pattern Lock. 

\section{Discussion}
\label{sec:tapmein_discussion}
Although results presented in our evaluation are encouraging, further investigation is needed as the data set is somewhat limited. Nevertheless, as a proof-of-concept, we have shown that \tapmein is potential for smartwatch authentication. Results can be seen as comparable to the state-of-the-art in the literature. We plan to conduct a real-world study to collect a data set which is not only larger but more realistic. 

Another limitation of the work is that we did not have enough data to analyze the effect of other conditions on the performance of TapMeIn, e.g., age or gender. For example, older users may have lower motor precision compared to young users, which may affect their recall of tap-passwords. This is a popular issue in touch-based research. We leave this topic for future work.

In the informal interview with users and attackers, we received some interesting feedback. Most users felt tapping is a natural task, just as how they tap along when listening to music, and they enjoyed \tapmein better. For the attackers, they perceived \tapmein was the hardest to attack among the three methods. Their strategy for video attack was to tap along when watching the video. Still, they found it difficult to replicate the victim's tap-password, especially when taking tap pressure and finger size into consideration. Interestingly, one attacker correctly recognized that a user tapped ``SOS'' in Morse code as her password. Despite that, FPR of this user was 0\%.  

Recently, acoustic side channel attacks on tap-based authentication have been introduced, which include human attack as well as automated attack based on signal processing~\cite{anand2015bad}. The idea was to extract a tap-password from tapping-sounds. The automated attack achieved an average accuracy of 85\%. However, these attacks may only achieve high success in systems that use only durations to represent a tap-password, like those in~\cite{wobbrock2009tapsongs, marques2013under, lin2011rhythmlink}. We argue that these attacks will not achieve much success in systems that utilize  behavioral and physiological characteristics as a second authentication factor, like \tapmein. As we have shown in our video attack evaluation, the attacker's success rate was only 4.1\%. 

Thermal attacks~\cite{abdelrahman2017stay}, which use thermal camera to capture heat traces resulting
from authentication, can be used to reconstruct PIN/patterns with high success rate (72\% to 100\% when performed within the first 30 seconds after authentication). We argue that \tapmein is not vulnerable to this attack. First, we observe that users tend to tap at a single spot on the screen which makes it really hard for thermal attack to infer how many taps are there in the password. Second, even knowing the number of taps, the attacker still needs to imitate tapping behavior of the user, which has been shown to be very difficult in our security evaluation (only 2.3\% FPR).

Regarding the comparative study between \tapmein and PIN and Pattern Lock, we found some interesting insights. First, in terms of error rate, PIN authentication on smartwatches produced an error rate that was more than twice as high as that of smartphones (average: 8.1\% compared to 3.1\%~\cite{harbach2016anatomy}). However, the error rate of Pattern Lock on smartwatches was better (average: 9.35\% compared to 12.1\% on smartphones~\cite{harbach2016anatomy}). This indicates fat-finger may be a problem for PIN method on smartwatches. Moreover, the error rate when entering patterns is four times as high as that when entering PINs on smartphones. Surprisingly, the results in our study suggest that on smartwatches, error rates are very similar between PIN and Pattern Lock. In terms of input time, PIN (average: 2180 ms) and Pattern Lock (1150 ms) on smartwatches were slower than PIN (average: 1963 ms) and Pattern (average: 910 ms) on smartphones \cite{harbach2016anatomy}. Note that our study was conducted in a lab whereas Harbach et al. conducted a field one. Nevertheless, these are very surprising usability results which deserve further investigation. We leave this for a longitudinal field study in future work.

\section{Related Work}
\label{sec:tapmein_related_work}
\subsection{Smartwatches and Wearables Authentication}
In recent years, authentication for smartwatches and wearables have been attracting attention from researchers. Bianchi et al. surveyed recent advances in the wearables authentication research and predicted that this topic will develop rapidly in the next decade~\cite{bianchi2016wearable}. Smartwatches and wearables possess limited input channels, i.e., small screen makes entering passwords laborious or difficult. However, wearables often have rich sensing capabilities. Moreover, with the advantage of being adorned by a user, wearables can be used to collect rich biometric information for explicit and implicit authentication. This is why a significant body of wearable authentication work has concentrated on building a behavioral model using sensor data~\cite{johnston2015smartwatch, kumar2016authenticating, shrestha2016walk, al2017unobtrusive, nguyen2017smartwatches}. 

For example, Johnston et. al proposed to build a gait biometric model for a user using data recorded by accelerometers and gyroscopes on smartwatch~\cite{johnston2015smartwatch}. They achieved 1.4\% EER with a 10-second window of sensor data for each authentication session. Similarly, Kumar et. al built a behavioral biometric of user's arm movement using accelerometers and gyroscopes sensor data on a smartwatch while a user was walking \cite{ kumar2016authenticating}. They evaluated multiple classifiers for authentication of user on a smartwatch. They reported best FPR of 2.2\% and FRR of 4.2\% with k-Nearest Neighbor classifier using a fusion feature set of both sensors. 

These kind of approaches have a drawback as they require a  user to walk, and for a period of time (i.e., 10 seconds) before an authentication decision can be made. This renders them only suitable for implicit authentication. Nassi et. al proposed  handwritten signature based user verification for wearables. The user wears a smartwatch or a wearable device while signing her signature on a paper. Sensors data captured from the signing process is used to authenticate the user. They reported an EER of 5\%. Our work is potentially the first touchscreen based authentication approach for smartwatches. We have shown that the fat-finger problem can be mitigated and  comparable performance to these implicit methods can be achieved.

\subsection{Tap-based Authentication}
Several tap-based authentication approaches have been introduced in the literature. As presented in the performance comparison section, most of them only used tap durations as features for authentication, i.e., TapSongs by Wobbrock et al. \cite{wobbrock2009tapsongs}, RhythmLink ~\cite{lin2011rhythmlink} or proposed approach by Marques et. al \cite{marques2013under}. Such features could not fully capture user's behavioral characteristics such as tap pressure and user's physiological characteristics like finger tip size. This was shown in either low accuracy or high false positive and false negative rates reported in these works.
RhyAuth, proposed by Chen et al. \cite{chen2015your}, is a rhythmic-based user authentication for mobile devices, which utilized behavioral and physiological features in user's tap as a second authentication factor. There are key differences between RhyAuth and \tapmein. First, RhyAuth assumed that negative samples are preloaded or downloaded from a server at training time, which, as discussed in Section~\ref{sec:tapmein_system} (System Design), does not work because totally random negative samples bias user's classifier, resulting in a poor classification performance when classifying real imitating samples. Second, the classifier of a user was trained and tested with samples from other users in their evaluation while \tapmein can be trained with only user's enrollment and synthesized samples, and tested with her login and imitating samples from attackers, which were ``unseen'' by the classifier. Thus, our results potentially reflect true performance in a real-world scenario. In addition, \tapmein is the first tap-based authentication evaluated for smartwatches.

Beside the application for authentication, tap and rhythmic patterns have also been proposed for improving user input and interaction with smartwatches. Ghomi et. al \cite{ghomi2012using} introduced Rhythmic Interaction. It allowed users to build a vocabulary of \emph{rhythmic patterns}, which were map to specific commands to form new input modality for small screen devices like smartwatches. Similarly, Oakley et. al \cite{oakley2015beats} proposed \emph{beat gestures} to extend input methods on smartwatches. Our work is different from these approaches in the sense that not only \tapmein needs to recognize tap pattern, it also needs to verify whether the pattern belongs to a specific user. This, in a way, is a more difficult task.

\section{Conclusion}
\label{sec:tapmein_conclusion}
This paper introduces \tapmein, a fast, accurate and secure two-factor authentication for smartwatches that allows users to tap a memorable melody anywhere on the screen to authenticate. Experimental results showed that \tapmein provides protection against shoulder surfing and video attacks, while maintaining short authentication time and low error rate. Comprehensive experiments were conducted to compare it with the de-facto PIN and Pattern Lock on smartwatches. Lastly, a user study rated \tapmein as highly usable and was favored for authentication in situations where a higher level of security is needed.

For future work, we plan to offer \tapmein as a real app on the app store to collect more data. Additionally, we would like to introduce \tapmein to people with visual impairments and evaluate its performance. We also want to study the change in tapping behavior of users over time and to explore classifier adapting/updating strategies for \tapmein, making it more effective in a large scale deployment. 

\section*{Acknowledgments}

We thank all participants for their time and insightful feedback. We thank anonymous reviewers for their constructive comments.

\balance{}

\bibliographystyle{SIGCHI-Reference-Format}

\end{document}